\documentclass[prb,twocolumn,showpacs,preprintnumbers,amsmath,amssymb]{revtex4}

\usepackage{graphicx}
\usepackage{dcolumn}
\usepackage{bm}
\usepackage{eucal}

\begin{document}
\title{Bose--Einstein condensation and the magnetically ordered
state of TlCuCl$_3$}
\author{Jens Jensen and Henrik Smith}
\affiliation{Niels Bohr Institute, Universitetsparken 5, 2100
Copenhagen, Denmark}
\date{July 3, 2009}

\begin{abstract} The dimerized $S=\frac{1}{2}$ spins of the Cu$^{2+}$ ions in
TlCuCl$_3$ are ordered antiferromagnetically in the presence of a
field larger than about 54 kOe in the zero-temperature limit. Within
the mean-field approximation all thermal effects are frozen out below
6 K. Nevertheless, experiments show significant changes of the
critical field and the magnetization below this temperature, which
reflect the presence of low-energetic dimer-spin excitations. We
calculate the dimer-spin correlation functions within a
self-consistent random-phase approximation, using as input the
effective exchange coupling parameters obtained from the measured
excitation spectra. The calculated critical field and magnetization
curves exhibit the main features of those measured experimentally,
but differ in important respects from the predictions of simplified
boson models.
\end{abstract}
 \maketitle
\section{Introduction}\label{sec:level1}
The concept of Bose--Einstein condensation dates back more than 80
years to the prediction of Einstein, based on Bose's work on the
statistics of photons, that a gas of non-interacting massive bosons
would condense below a certain critical temperature $T_c$. The
condensation implies that below $T_c$ a non-zero fraction of the
total number of particles occupies the lowest single-particle quantum
state. For dilute atomic gases this phenomenon was realized
experimentally in 1995 for trapped clouds of alkali atoms (see e.g.\
Ref.\ \onlinecite{HS}).

For a uniform gas of density $n$ the transition temperature is given
by $kT_c^{}\approx 3.31\hbar_{}^2n_{}^{2/3}/m$, where $m$ is the
particle mass. For particles trapped in a harmonic oscillator
potential (trap frequencies $\omega_x^{}$, $\omega_y^{}$ and
$\omega_z^{}$) one has $kT_c^{}\approx 0.94 \hbar
(N\omega_x^{}\omega_y^{}\omega_z^{})^{1/3}$, where $N$ is the total
number of particles. In the latter case the particle mass enters
through the trap frequencies, equal to the square root of the force
constants in the three directions divided by the particle mass. When
a trapped gas is dilute in the sense that the atom--atom scattering
length is much less than the interatomic distance, the observed
transition temperatures agree well with theoretical expectation for a
non-interacting gas. For less dilute gases interaction effects give
rise to an observable small shift of $T_c^{}$ proportional to the
scattering length.

The condensation of massive bosons into a single quantum state is
intimately connected to the conservation of particle number. For
massless bosons such as phonons or magnons the particle number is not
fixed but depends on temperature, and there is therefore no
Bose--Einstein condensation in the traditional sense of the term.
However, there has been a wide use of model Hamiltonians for magnetic
systems that have features in common with those of interacting,
massive bosons. The aim of the present work is to consider one such
specific system, that of the dimerized Cu$^{2+}$ spins in TlCuCl$_3$,
and compare predictions of such models with calculations that are
based on (approximate) solutions of the many-body problem of
interacting spins. A recent review of experimental and theoretical
developments concerning the magnetic ordering of TlCuCl$_3$ and
related compounds has been given by Giamarchi {\it et
al.}\cite{Giamarchi}

Magnetization measurements\cite{Oosa99,Raffaele} and inelastic
neutron-scattering experiments\cite{Cavadini,Oosawa} demonstrate
clearly that nearest-neighboring pairs of $S=\frac{1}{2}$ spins of
the Cu$^{2+}$ ions in TlCuCl$_3$ are dimerized leading to an $S=0$
ground state and an $S=1$ excited triplet around 5.2--5.7 meV above
the singlet. Due to the exchange interactions between the dimers the
excitations become strongly dispersive, and, in the zero-temperature
limit, the minimum energy of the degenerate singlet-triplet mode, at
(001), is only about 0.7 meV. When a field is applied, the energy of
one of the three normal modes is reduced and goes to zero at a
critical field of about 54 kOe at zero temperature. The Cu spins of
isostructural KCuCl$_3$ are similarly dimerized, but the interdimer
interactions are relatively weaker and the critical field is about
230 kOe at $T=0$ in this system.\cite{OosawaK} The phase transition
shown by TlCuCl$_3$ at the field where the excitation energy
vanishes, has been analyzed by Nikuni {\it et al.}\cite{Nikuni} They
assumed the dimer system to be described by an effective Haniltonian
of the form
\begin{equation}
{\cal H} = \sum_{{\bm k}}\left(\frac{\hbar^2k^2}{2m}-\mu\right)
a_{{\bm k}}^{\dagger}a_{{\bm k}}^{\phantom{\dagger}}
+\frac{v_0}{2}\sum_{{\bm k},{\bm k'}\!, {\bm q}} a_{{\bm k}+{\bm
q}}^{\dagger} a_{{\bm k'}-{\bm q}}^{\dagger} a_{{\bm
k'}}^{\phantom{\dagger}}a_{{\bm k}}^{\phantom{\dagger}}\,, \label{e0}
\end{equation}
where the bosonic operators $a^{\dagger}$ and $a$ denote ``magnon''
creation and annihilation operators and the positive constant $v_0$
denotes the strength of the repulsive magnon--magnon interaction,
assumed to be a delta function in real space. The quantity $\mu$
plays the role of a chemical potential, assumed proportional to the
difference between the applied magnetic field and the critical field.
Using $v_0$ and $m$ as fitting parameters, Nikuni {\it et al.}\ were
able to give a reasonable account of the temperature dependence of
the critical field and the magnetization along the applied field in
the ordered antiferromagnetic  state. An extended version of their
theory based on a more realistic dispersion of the magnetic
excitations was presented by Misguich and Oshikawa.\cite{Misguich}
However, as we shall see in detail in Sec.\ IV below, these
simplified boson models suffer from inconsistencies that originate in
their neglect of the highest level of the triplet. Another, more
general problem with boson models is that double occupancy of a local
site should be prohibited. In this connection we mention the work of
Sirker {\it et al.},\cite{Sirker} who used a bond-operator approach
to map the spin system onto a model of interacting bosons by
introducing an infinite on-site repulsion between local triplet
excitations.

In the following treatment of the dimerized spin system in TlCuCl$_3$
we adopt a different point of view and start from the Hamiltonian for
the spin system itself, using as input the effective exchange
coupling that has been derived from measured excitation spectra. Our
approach is a generalization of the zero-temperature theory by
Matsumoto {\it et al.},\cite{Matsumoto} who used the random-phase
approximation (RPA) to calculate magnetization curves and excitation
spectra for TlCuCl$_3$. They also considered the case when the phase
transition is induced by the application of a hydrostatic
pressure.\cite{Ruegg2,Ruegg08} Here we only address the case of a
field-induced transition, but the theory of Matsumoto {\it et al.} is
extended to include both the effects of quantum fluctuations at zero
temperature and the effects of thermal fluctuations. The
self-consistent version of the RPA, which is the one applied here, is
faced with the similar problem of double occupancy as the boson
modelling. However, here this problem is found to have a natural
solution by a consideration of the higher order modifications of the
Green functions. The self-consistent RPA theory for the paramagnetic
phase of the dimer-spin system is presented in Sec.\ II, which, in
Sec.\ III, is followed by an analysis of the antiferromagnetic phase.
A closer examination and discussion of the results obtained are
referred to the last Sec.\ IV.

\section{Excitations in the paramagnetic phase}\label{sec:level2}
\subsection{The self-consistent RPA theory}
The TlCuCl$_3$ crystal is monoclinic (space group $P2_1/c$) and the
lattice parameters are $a=3.9815$ \AA, $b=14.144$ \AA, $c=8.8904$
\AA\ and $\beta=96.32^\circ$ at room temperature.\cite{Tanaka} The
crystal is constructed from layers with configuration Cu$_2$Cl$_6$
stacked on top of each other so as to form two chains of Cu ions
parallel to the $a$ axis. The chains are separated by Tl ions and
pass through the center and corners of the $b$--$c$ plane in the unit
cell. There are four Cu ions or two dimer pairs per unit cell. The
dimer pair in the unit cell belonging to the chain through a corner
is located at site 1: $(x,y,z)$ and site 2:
$(\bar{x},\bar{y},\bar{z})$, and the pair belonging to the other
chain is placed at site 3: $(x,\bar{y}+\frac{1}{2},z+\frac{1}{2})$
and site 4: $(\bar{x},y+\frac{1}{2},\bar{z}+\frac{1}{2})$. Here
$x=0.2338$, $y=0.0486$, and $z=-0.0175$, and the numbering of the
sites from 1 to 4 defines the four different Cu-sublattices.

The Hamiltonian is assumed to be
\begin{equation}
\label{e1}{\cal H}=-{\textstyle\frac{1}{2}}\sum_{ij}{\cal J}(ij){\bm
s}_i^{}\cdot{\bm s}_j^{}-g\mu_B^{}\sum_i{\bm H}\cdot{\bm s}_i^{}\,,
\end{equation}
where ${\bm s}_i^{}$ is the spin-variable of the Cu ion at the $i$th
site. The most important exchange parameter is $\Delta=-{\cal
J}(i_1i_2)$, where $(i_1i_2)$ are the nearest-neighbor Cu pairs (the
1-2 or the 3-4 ions in the unit cell). The Fourier transform of the
Heisenberg exchange interactions between spins on sublattice $\alpha$
and $\beta$ is defined in terms of the remaining coupling parameters
\begin{equation}
\label{e2}{\cal J}_{\alpha\beta}^{}({\bm q})=\!\!\sum_{j\in
\beta\text{-subl.}} \hspace{-10pt}'\hspace{5pt}{\cal
J}(ij)\,e^{-i{\bm q}\cdot({\bm R}_i-{\bm R}_j)},\quad
i\in~\alpha\mbox{-subl.},
\end{equation}
where ${\bm R}_i^{}$ is the position of the $i$th dimer, and the
prime indicates that the dominating interaction $\Delta$ is excluded
from the sum, $(ij)\ne(i_1i_2)$.

When the interactions between the dimers are neglected, the
Hamiltonian may be diagonalized exactly in terms of independent
products of single-dimer eigenstates. The total spin of the $i$th
dimer is ${\bm S}_i^{}={\bm s}_{i_1}+{\bm s}_{i_2}$, where ${\bm
s}_{i_1}$ and ${\bm s}_{i_2}$ denote the Cu spins belonging to,
respectively, the sublattices 1 and 2, or 3 and 4. The total spin
defines the basis $|SS_z\rangle$, and when the field is along the $z$
axis, the eigenstates of the non-interacting dimer are
\begin{itemize}
\item[]state $|3\rangle=|1-1\rangle$ at the energy $\Delta+h$,
\item[]state $|2\rangle=|10\rangle$ at the energy $\Delta$,
\item[]state $|1\rangle=|1+1\rangle$ at the energy $\Delta-h$,
\item[]state $|0\rangle=|00\rangle$ at zero energy,
\end{itemize}
with $h=g\mu_B^{}|{\bm H}|$ and $\Delta$ positive. When $h<\Delta$
the ground state is the non-magnetic singlet $|00\rangle$. In the
present section we focus on this condition, and we shall assume that
the system stays paramagnetic also in the presence of the interdimer
interaction ${\cal J}_{\alpha\beta}^{}({\bm q})$. The original
Hamiltonian (\ref{e1}) may be rewritten in terms of two dimer-spin
variables, the sum ${\bm S}_i={\bm s}_{i_1}+{\bm s}_{i_2}$ and the
difference $\bar{\bm S}_i={\bm s}_{i_1}-{\bm s}_{i_2}$.  The sum
operator only has non-zero matrix elements between the three excited
$S=1$ states. We consider the case where the populations of these
levels are small (at sufficiently low temperatures in the disordered
phase), in which case the dynamical effects due to ${\bm S}_i$ are
negligible. When ${\bm S}_i$ is neglected, the Hamiltonian involves
only a single effective ${\bm q}$-dependent exchange term
$-\frac{1}{2}\sum_{\bm q} J({\bm q})\bar{\bm S}_{\bm q}\cdot\bar{\bm
S}_{-\bm q}$ with
\begin{eqnarray}
\label{e3}J({\bm q})&=&\frac{1}{4}\left[{\cal J}_{11}({\bm q})+{\cal
J}_{22}({\bm q})-{\cal J}_{12}({\bm q})-{\cal J}_{21}({\bm
q})\right]\nonumber\\ &\pm&\frac{1}{4}\left[{\cal J}_{13}({\bm
q})+{\cal J}_{24}({\bm q})-{\cal J}_{14}({\bm q})-{\cal J}_{23}({\bm
q})\right]\,.
\end{eqnarray}
The presence of two equivalent dimers per unit cell yields two values
for the effective interaction for each value of the wave vector
within the first Brillouin zone. Alternatively, one may use an
extended zone scheme with an effective basis of one dimer per unit
cell, in which case only the upper sign applies.

In order to study the spin dynamics of this Hamiltonian we introduce
the standard basis operators\cite{Haley,Bak,JJ} for the $j$th dimer
\begin{equation}
\label{e4} a_{\mu\nu}^j=\left(|\mu\rangle\langle\nu|\right)_j\,,\quad
\mu,\,\nu = 0,\,1,\,2,\,3\,.
\end{equation}
In the present case of a dimer system with stationary bonds, these
operators serve the same purpose but are of more general use than the
``bond operators'' applied by Matsumoto {\it et al.}\cite{Matsumoto}
In terms of the standard basis operators the components of $\bar{\bm
S}_j$ become
\begin{eqnarray}
\label{e5}
&\bar{S}_{jx}^{}=&\frac{1}{\sqrt{2}}\left[a_{30}^j-a_{10}^j+
a_{03}^j-a_{01}^j\right],\cr
&\bar{S}_{jy}^{}=&\frac{i}{\sqrt{2}}\left[a_{30}^j+a_{10}^j-
a_{03}^j-a_{01}^j\right],\cr &\bar{S}_{jz}^{}=&a_{20}^j+a_{02}^j\,,
\end{eqnarray}
and the Hamiltonian may be written
\begin{eqnarray}
\label{e6} {\cal
H}=\sum_i^{}\left[(\Delta-h)\,a_{11}^i+\Delta\,a_{22}^i+
(\Delta+h)\,a_{33}^i\right]\qquad&&\cr-\sum_{ij}
J(ij)\left[a_{01}^ia_{10}^j
+a_{03}^ia_{30}^j-a_{01}^ia_{03}^j-a_{10}^ia_{30}^j\right.&&\cr
+\left.a_{02}^ia_{20}^j+{\textstyle\frac{1}{2}}
\left(a_{20}^ia_{20}^j+a_{02}^ia_{02}^j\right)\right],&&
\end{eqnarray}
when only the Fourier transform $J(ij)$ of the effective interaction,
Eq.\ (\ref{e3}), is included. Next we define a $6\times6$ matrix of
Green functions\cite{Zubarev}
\begin{equation}
\label{e7}
\bar{\bar{G}}(ij,\omega)=-\frac{i}{\hbar}\int_{-\infty}^\infty
\!\!\theta(t)\big\langle\big[{\bm a}_i^{}(t),{\bm
a}_j^\dagger(0)\big]\big\rangle e^{i\omega t}dt\,.
\end{equation}
A single bracket $\langle\cdots\rangle$ denotes the thermal
expectation value, and ${\bm a}_i^{}(t)$ is a vector operator of site
$i$ at time $t$ with components
\begin{equation}
\label{e8} {\bm a}_i^{}=\left(a_{01}^i,a_{10}^i,a_{02}^i,a_{20}^i,
a_{03}^i,a_{30}^i\right).
\end{equation}
With the short-hand double-bracket notation
$\bar{\bar{G}}(ij,\omega)=\langle\langle{\bm a}_i^{};{\bm
a}_j^\dagger\rangle\rangle$, the equations of motion for the Green
functions are (see for instance Ref.\ \onlinecite{JJ})
\begin{equation}
\label{e9} \hbar\omega\big\langle\big\langle{\bm a}_i^{};{\bm
a}_j^\dagger\big\rangle\big\rangle-\big\langle\big\langle\big[{\bm
a}_i^{},{\cal H}\big];{\bm
a}_j^\dagger\big\rangle\big\rangle=\big\langle\big[{\bm a}_i^{},{\bm
a}_j^\dagger\big]\big\rangle\,,
\end{equation}
where the new higher-order Green functions introduced by the second
term are determined by the Hamiltonian (\ref{e6}) by  the use of the
commutator relation
\begin{equation}
\label{e10}
\big[a_{\mu\nu}^i,a_{\mu'\nu'}^j\big]=\delta_{ij}^{}\left(
\delta_{\nu\mu'}^{}
a_{\mu\nu'}^{i}-\delta_{\mu\nu'}^{}a_{\mu'\nu}^{i}\right)\,.
\end{equation}
By using an RPA decoupling of the higher-order Green functions,
$a_{\mu\nu}^ia_{\mu'\nu'}^j\approx a_{\mu\nu}^i\langle
a_{\mu'\nu'}^j\rangle+\langle a_{\mu\nu}^i\rangle a_{\mu'\nu'}^j$,
one finds that the equations of motion are reduced to a closed set of
equations, which are solvable by a Fourier transformation. Further,
the $6\times6$ matrix equations decouple into 3 sets of $2\times2$
matrix equations, and one of these is
\begin{equation}
\label{e11}
\left(\begin{matrix}\hbar\omega-E_1^{}& -J_1^{}\\
J_3^{} & \hbar\omega+E_3^{}\end{matrix}\right)
\left(\begin{matrix}G_{11}^{} & G_{16}^{}\\
G_{61}^{} & G_{66}^{}\end{matrix}\right)=
\left(\begin{matrix}n_{01}^{} & 0\\
0 & -n_{03}^{}\end{matrix}\right).
\end{equation}
The Green functions depend on the Fourier variables,
$G_{\mu\nu}^{}=G_{\mu\nu}^{}({\bm q},\omega)$, and we have introduced
the following parameters
\begin{eqnarray}
\label{e12} E_1^{}=\Delta-h-J_1^{}\,,\qquad J_1^{}=n_{01}^{}J({\bm
q})\,,\nonumber\\ E_3^{}=\Delta+h-J_3^{}\,,\qquad
J_3^{}=n_{03}^{}J({\bm q})\,,
\end{eqnarray}
with $n_{\mu}^{}$ being the average population of the $\mu$th dimer
level
\begin{equation}
\label{e13} n_{\mu}^{}=\langle a_{\mu\mu}^i\rangle\,,\quad\mbox{and
the difference}\quad n_{\mu\nu}=n_\mu^{}-n_\nu^{}\,.
\end{equation}
Inversion of the first matrix in Eq.\ (\ref{e11}) results in
\begin{eqnarray}
\label{e14}\left(\begin{matrix}G_{11}^{} &
G_{16}^{}\\
G_{61}^{} & G_{66}^{}\end{matrix}\right)&=&\frac{-1}{(E_{\bm
q}^--\hbar\omega)(E_{\bm q}^++\hbar\omega)}\\
&\times&\left(\begin{matrix}n_{01}^{}(E_3^{}+\hbar\omega) &
-n_{03}^{}J_1^{}\\ -n_{01}^{}J_3^{} &
n_{03}^{}(E_1^{}-\hbar\omega)\end{matrix}\right)\,.\nonumber
\end{eqnarray}
The poles of the Green functions determine the excitation energies
$E_{\bm q}^\pm$, which are given by
\begin{eqnarray}
\label{e15} &&E_{\bm q}^\pm=E_{\bm
q}^{}\pm\left(h-{\textstyle\frac{1}{2}}
n_{13}^{}J({\bm q})\right),\\
&&E_{\bm q}^2=\Delta^2-(n_{01}^{}+n_{03}^{})\Delta J({\bm q})+
\left({\textstyle\frac{1}{2}} n_{13}^{}J({\bm q})\right)^2.\nonumber
\end{eqnarray}
The (2,5) part of $\bar{\bar{G}}({\bm q},\omega)$ is given by the
same expression as the (1,6) part in Eq.\ (\ref{e14}), except that
$\omega$ is replaced by $-\omega$. The result for the (3,4) part is
that obtained from Eq.\ (\ref{e14}) for $h=0$, corresponding to the
replacement of $n_{01}^{}$ and $n_{03}^{}$ by $n_{02}^{}$, and it
leads to poles at the energies $\pm E_{\bm q}^z$, where
\begin{equation}
\label{e16} E_{\bm q}^z=\sqrt{\Delta^2-2\Delta J_2^{}}\,,\quad
J_2^{}=n_{02}^{}J({\bm q})\,.
\end{equation}

We introduce the following matrix of equal-time correlation functions
\begin{equation}
\label{e17} \bar{\bar{A}}({\bm q}^{})=\frac{1}{N}\sum_{ij}
\big\langle{\bm a}_i^{}\,{\bm a}_j^\dagger\big\rangle \,e^{-i{\bm
q}\cdot({\bm R}_i-{\bm R}_j)}\,,
\end{equation}
where $N$ is the number of dimers. According to the
fluctuation-dissipation theorem (see e.g.\ Ref.\ \onlinecite{JJ})
\begin{equation}
\label{e18}\bar{\bar{A}}({\bm
q}^{})=-\frac{1}{\pi}\int_{-\infty}^\infty
\frac{1}{1-e^{-\beta\hbar\omega}}\,\bar{\bar{G}}''({\bm
q},\omega)\,d(\hbar\omega)\,,
\end{equation}
where $\bar{\bar{G}}''$ denotes the imaginary part of the matrix
Green function and $\beta=1/kT$. By definition, the average value of,
for instance, the $11$-component is
\begin{equation}
\label{e19} \overline{A}_{11}^{}=\frac{1}{N}\sum_{\bm
q}A_{11}^{}({\bm q})=\langle a_{01}^i\,a_{10}^i\rangle=\langle
a_{00}^i\rangle=n_0^{}\,.
\end{equation}
Hence, by calculating the ${\bm q}$ averages of the correlation
functions, the Green functions may be used for determining the
populations of the dimer levels. From the diagonal part of the
$\overline{A}$-matrix, we get straightforwardly
\begin{eqnarray}
\label{e20} \frac{n_0+n_1}{2n_{01}}+\frac{n_0+n_3}{2n_{03}}\!&=&\!
\frac{1}{N}\sum_{\bm q}\frac{E_1^{}+E_3^{}}{E_{\bm q}^++ E_{\bm
q}^-}\left(1+n_{\bm q}^-+n_{\bm q}^+\right),\nonumber\\
\frac{n_0+n_1}{2n_{01}}-\frac{n_0+n_3}{2n_{03}}\!&=&\!
\frac{1}{N}\sum_{\bm q}\left(n_{\bm q}^--n_{\bm q}^+\right)
\end{eqnarray}
and
\begin{equation}
\label{e21}\frac{n_0+n_2}{n_{02}}=\frac{1}{N}\sum_{\bm
q}\frac{\Delta-J_2^{}}{E_{\bm q}^z}\left(1+2n_{\bm q}^z\right)\,,
\end{equation}
where
\begin{equation}
\label{e22} n_{\bm q}^\pm=\frac{1}{e^{\beta E_{\bm
q}^\pm}-1}\,,\qquad n_{\bm q}^z=\frac{1}{e^{\beta E_{\bm q}^z}-1}\,.
\end{equation}
The three equations determine the four population numbers, when they
are supplemented by the exact condition that
\begin{equation}
\label{e23} n_0^{}+n_1^{}+n_2^{}+n_3^{}=1\,.
\end{equation}
The RPA decoupling is valid in the mean-field (MF) approximation,
where the thermal averages are determined by the Hamiltonian ${\cal
H}_{\text{MF}}^{}$ for the non-interacting system, i.e.\ within the
approximation $\langle a_{\mu\nu}^{}\rangle\approx\mbox{Tr}[
a_{\mu\nu}^{}\exp(-\beta{\cal
H}_{\text{MF}}^{})]=\delta_{\mu\nu}^{}n_\mu^{\text{MF}}$. In the case
of strong dispersion, this approximation certainly underestimates the
populations of the excited levels. Anticipating that the correlation
effects predicted by the RPA theory are reasonably trustworthy, the
self-consistent equations above should lead to a more accurate
determination of the population numbers than that offered by the MF
approximation. Unfortunately, the present theory also predicts that
the off-diagonal components of the $\overline{A}$ matrix are non-zero
contradicting that, for instance, $\overline{A}_{16}^{}=\langle
a_{01}^ia_{03}^i\rangle$ should vanish identically. Phrased
differently, this inconsistency implies that the RPA result for the
occupation numbers is not unique but depends on the particular choice
of correlation functions used in the calculation. A non-zero value of
$\langle a_{01}^ia_{03}^i\rangle$ is the equivalent of a double
occupancy of bosons at a single site. Instead of introducing an
arbitrary repulsive potential, we are here going to consider possible
improvements of the RPA-decoupling procedure applied above.

The present system is in many ways similar to the singlet-doublet
system encountered in praseodymium metal. Improvements of the RPA for
this system have been derived in Ref.\ \onlinecite{JJPr}, and this
theory has been applied to the calculation of the linewidths and the
energy renormalization of the excitations in two other dimer systems
Cs$_2$Cr$_2$Br$_9$ and KCuCl$_3$ for the case of zero
field.\cite{Leu,CavadiniK} It is straightforward to generalize the
singlet-doublet theory so as to account for the presence of a third
level. Here, we are going to extend the theory to the case where the
field is non-zero, and, in the next section, to consider the
modifications produced by an ordered moment. The presence of the
interdimer interactions $J(ij)$ implies that $a_{00}^i$ does not
commute with the Hamiltonian. Hence, the assumed ground state, the
product state of $|0\rangle_i $, is not an eigenstate of the
interacting system. The situation compares with the simple Heisenberg
antiferromagnet, where the mean-field N\'eel state is not the true
ground state. In this case the RPA theory predicts a zero-temperature
reduction of the antiferromagnetic moment from its saturated N\'eel
state value. Equivalently, the RPA results above imply that $n_0^{}$
is smaller than its saturation value 1 at zero temperature. The
single-dimer population numbers are subject to quantum as well as
thermal fluctuations.

One of the terms neglected in the RPA equation (\ref{e11}) involves
the Green function $\langle\langle(a_{00}^i-a_{11}^i-
n_{01}^{})a_{01}^j;{\bf a}_k^\dagger\rangle\rangle$. Since $a_{00}^i$
and $a_{11}^i$ are not true constants of motion, this Green function
may modify the RPA result. The equations of motion for the Green
functions neglected in Eq.\ (\ref{e11}) have been analyzed in Ref.\
\onlinecite{JJPr}. The consequences are that the RPA parameters are
being replaced by effective ones and that the excitations become
damped. One of the effective parameters $J_{xy}^{}({\bm q})$ replaces
$J({\bm q})$ in $J_1^{}$ and $J_3^{}$ in all of the equations above
and is\cite{JJPr}
\begin{equation}
\label{e24} J_{xy}^{}({\bm q})=J({\bm q})-a_{xy}^{}+
\eta_{xy}^{}\left[b_{xy}^{}(\omega)-a_{xy}^{}\right]\,,
\end{equation}
where
\begin{equation}
\label{e25} \eta_{xy}^{}=\left(\frac{2}{n_{01}+n_{03}}\right)^2-1\,.
\end{equation}
The most important modification of $J({\bm q})$ is the constant shift
introduced by $a_{xy}^{}$, but first we want to discuss the other
term
\begin{equation}
\label{e26} b_{xy}^{}(\omega)=\frac{1}{N}\sum_{\bm k}\left[J({\bm
k})\right]^2\chi_{xy}^{}({\bm k,\omega})\,,
\end{equation}
where $\chi_{xy}^{}({\bm k,\omega})$ is a generalized
susceptibility.\cite{JJPr} The imaginary part of $b_{xy}^{}(\omega)$
determines the damping effects, which are, however, small at low
temperatures. The renormalization of $J_{xy}^{}({\bm q})$ produced by
the real part of $b_{xy}^{}(\omega)$ is somewhat smaller than the
constant shift due to $a_{xy}^{}$, but it is not entirely negligible.
Because of its moderate importance we have simplified the expression
for the $b_{xy}^{}(\omega)$ term as follows
\begin{eqnarray}
\label{e27}b_{xy}^{}(\omega)&\approx& \mbox{Re}\left[b_{xy}^{}(E_{\bm
q}^{}/\hbar)\right]\approx
\left(\frac{n_{01}^{}+n_{03}^{}}{n_{01}^0+n_{03}^0}\right)^3
B_{\bm q}^{}\,,\nonumber\\
B_{\bm q}^{}&=& \frac{1}{N}\sum_{\bm k}\frac{[J({\bm k})]^2[J({\bm
q})-J({\bm k})]} {[J({\bm q})-J({\bm k})]^2+\epsilon_0^2}\,.\qquad
\end{eqnarray}
We include only the real part and neglect modifications produced by
the field. The quantity $\epsilon_0^{}$ is due to the finite lifetime
of the excitations and we take it to be a constant,
$\epsilon_0^{}\approx0.1$ meV at zero temperature and field. The
linewidths are going to increase rapidly when the temperature becomes
comparable to $\Delta/k$. We have accounted for this effect in a
rough manner by scaling the result $B_{\bm q}$ at zero temperature
and field by the population-dependent factor in front, where
$n_\mu^0$ is the value of a population number at zero field and
temperature. A closer examination indicates that this simple scaling
accounts for the increase of the linewidths in a reasonable way. The
scale factor is unimportant for the analysis of the low temperature
properties, but the power of 3 used in this expression ensures that
the scale factor times $\eta_{xy}$ vanishes in the high-temperature
limit.

Finally, the most important renormalization effect, the constant term
$(1+\eta_{xy})a_{xy}$ in $J_{xy}^{}({\bm q})$ in Eq.\ (\ref{e24}), is
determined implicitly by
\begin{equation}
\label{e28} \frac{1}{N}\sum_{\bm q}\frac{J_{xy}^{}({\bm q})}{E_{\bm
q}^++ E_{\bm q}^-}\left(1+n_{\bm q}^-+n_{\bm q}^+\right)=0\,.
\end{equation}
The renormalized value of $\overline{A}_{16}^{}$ is equal to this sum
over ${\bm q}$ times $n_{01}^{}n_{03}^{}$, and since the sum now
vanishes, the condition $\langle a_{01}^ia_{03}^i\rangle=0$ is
satisfied. The parameters $J_1^{}$ and $J_3^{}$ in Eqs.\
(\ref{e11})-(\ref{e15}) are replaced by, respectively,
$n_{01}^{}J_{xy}^{}({\bm q})$ and $n_{03}^{}J_{xy}^{}({\bm q})$, and,
similarly, $J_2^{}$ in Eq.\ (\ref{e16}) is replaced by
$n_{02}^{}J_z^{}({\bm q})$, where the effective exchange coupling is
\begin{equation}
\label{e29} J_z^{}({\bm q})=J({\bm q})-a_z^{}+
\eta_z^{}\left[b_z^{}(\omega)-a_z^{}\right]\,.
\end{equation}
Here $\eta_z^{}=1/n_{02}^2 -1$ and
$b_z^{}(\omega)\approx(n_{02}^{}/n_{02}^0)^3B_{\bm q}^{}$. The
constant term $(1+\eta_z)a_z$ is determined by the condition that
$\langle a_{02}^ia_{02}^i\rangle=0$. Besides the modifications of the
exchange couplings, the energy splitting $\Delta$ in $E_1^{}$ and
$E_3^{}$, or in $E_2^{}$, is replaced by, respectively,\cite{JJPr}
\begin{eqnarray}
\label{e30} &&\Delta_1^{}=\Delta_3^{}=\Delta
+(a_{xy}^{}+a_z^{})/2\,,\nonumber\\
&&\Delta_2^{}=\Delta+a_{xy}\,.
\end{eqnarray}
The renormalization of the energy-level separations has not much
influence on the final calculations, and the small additional
modifications derived in Ref.\ \onlinecite{JJPr} might have been
neglected. However, the leading order effects of these extra terms
are actually included above by assuming $1+\eta_{xy}^{}$ to be a
factor $n_0^{}+(n_1^{}+n_3^{})/2$ smaller than derived in Ref.\
\onlinecite{JJPr}, and, similarly, $1+\eta_{z}^{}$ has been divided
by $n_0^{}+n_2^{}$.

\begin{figure}[t]
\includegraphics[width=0.99\linewidth]{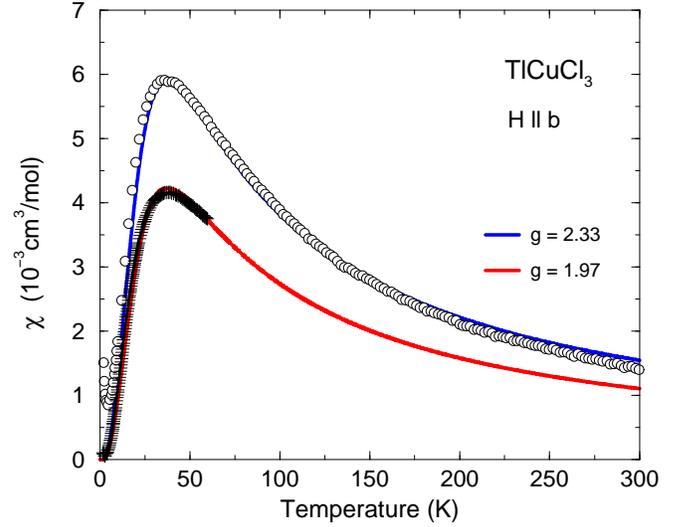}
\caption{(Color online) The susceptibility of TlCuCl$_3$ determined
experimentally with a field of 10 kOe along the $b$ axis. The pluses
show the data of Oosawa {\it et al.},\cite{Oosa99} and the open
circles are the results of Dell'Amore {\it et al.}\cite{Raffaele} The
experimental results are compared with the calculated ones obtained
by assuming $g=1.97$ or $g=2.33$. In all other calculations we use
$g=2.06$.} \label{f1}
\end{figure}

\subsection{Comparison with experiments}
Cavadini {\it et al.}\cite{Cavadini} and Oosawa {\it et
al.}\cite{Oosawa} have measured the dispersion of the magnetic
excitations in TlCuCl$_3$ at zero field in the zero-temperature limit
(1.5 K). The two sets of results agree where they overlap, and the
combined experimental results are most closely reproduced by the
dispersion parameters derived by Oosawa {\it et al.} The parameters
used here (in units of meV) determine the effective exchange coupling
according to
\begin{eqnarray}
\label{e31} J_{\text{eff}}^{}({\bm q})&=&0.46\cos({\bm q}\cdot{\bm
a}) -0.05\cos(2{\bm q}\cdot{\bm a})\nonumber\\&+&1.53\cos\left({\bm
q}\cdot(2{\bm a}+{\bm c})\right)\nonumber\\ &\mp& 0.86\cos\left({\bm
q}\cdot({\bm a}+{\textstyle\frac{1}{2}}{\bm
c})\right)\cos\left({\textstyle\frac{1}{2}}{\bm q}\cdot{\bm
b}\right)\,.
\end{eqnarray}
These are the parameters derived by Oosawa {\it et al.} except that
their coupling between the two chains\\ $\mp[0.98\cos({\bm
q}\cdot({\bm a}+\frac{1}{2}{\bm c}))-0.12 \cos(\frac{1}{2}{\bm
q}\cdot{\bm c})]\cos(\frac{1}{2}{\bm q}\cdot{\bm b})$ has been
approximated by a single term. The three modes are degenerate at zero
field, and the dispersion relation assumed by Oosawa {\it et al.}\ in
their analysis is $E_{\bm q}^2=\Delta_{\text{eff}}^2-
2\Delta_{\text{eff}}^{}J_{\text{eff}}^{}({\bm q})$ implying that
\begin{eqnarray}
\label{e32} &&\Delta_{\text{eff}}^2=\frac{1}{N}\sum_{\bm q}E_{\bf
q}^2=
(\Delta+a_0^{})\left(\Delta+a_0^{}+2a_0^{}/n_{01}^{0}\right),\nonumber\\
&&J_{\text{eff}}^{}({\bm q})=n_{01}^0\left[J({\bm
q})+\eta_{xy}^0B_{\bm
q}\right]\frac{\Delta+a_0^{}}{\Delta_{\text{eff}}}\,.
\end{eqnarray}
Here $a_0^{}$ is the value of $a_{xy}^{}$ or $a_z^{}$ at zero
temperature and field. The quantity $B_{\bm q}^{}$ is roughly
proportional to $J({\bm q})$ and its averaged value with respect to
${\bm q}$ is zero. The RPA equations above have been solved
numerically by an iterative procedure. The calculations benefit from
the fact that all ${\bm q}$ summations may be parameterized in terms
of $J_{\text{eff}}^{}({\bm q})$ [we neglect the minor difference
between $J({\bm q})+\eta_{xy}^0B_{\bm q}^{}$ and $J({\bm q})$ in Eq.\
(\ref{e27}) determining $B_{\bm q}^{}$]. Hence all summations may be
expressed as integrals with respect to $J_{\text{eff}}^{}({\bm q})$
times a corresponding ``density of states'' calculated once and for
all from Eq.\ (\ref{e31}). The value of $\Delta_{\text{eff}}^{}$ used
in the calculations is 5.671 meV, (almost) equal to the value of 5.68
meV derived by Oosawa {\it et al.}\cite{Oosawa} Besides this
parameter and those defining $J_{\text{eff}}^{}({\bm q})$ we have
assumed that $g=2.06$, which is the generally accepted value for $g$
in the case where the field is applied along the $b$
axis.\cite{Oosa99,Raffaele,Kolezhuk} Finally, we have added the mean
field from the parallel component to the applied field so that
$g\mu_B^{}H=h_0^{}$ in Eq.\ (\ref{e1}) is replaced by
$h=h_0^{}+J_F^{}({\bm0})\langle S_z^{}\rangle$. The ferromagnetic
coupling $J_F^{}({\bm0})$ is estimated to be about $-1.9$ meV by
Dell'Amore {\it et al.}\cite{Raffaele} and of the order of $-2.8$ meV
by Oosawa {\it et al.}\cite{Oosawa} (using their parameters
determined by a cluster series expansion). Here we assume
$J_F^{}({\bm0})=-2.4$ meV. The moment per Cu$^{2+}$ ion parallel to
the applied field is
\begin{equation}
\label{e33} m_z^{}=\frac{1}{N}\sum_i g\mu_B^{}\frac{1}{2}\langle
S_{iz}^{}\rangle=g\mu_B^{}\frac{n_1^{}-n_3^{}}{2}\,.
\end{equation}
It is worthwhile to notice that although the population numbers of
the excited levels are predicted to be non-zero at $T=0$, Eq.\
(\ref{e20})-(\ref{e23}), the quantum fluctuations do not give rise to
any difference between $n_1^{}$ and $n_3^{}$, i.e.\ $m_z^{}=0$ at
zero temperature as long as the system stays paramagnetic. This is
consistent with the condition that $\sum_i
S_{iz}^{}=\sum_i(a_{11}^i-a_{33}^i)$ commutes with the Hamiltonian.

\begin{figure}[t]
\includegraphics[width=0.99\linewidth]{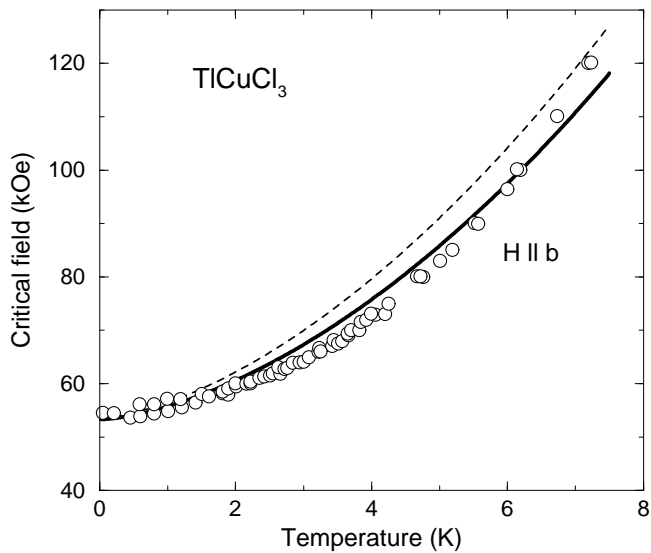}
\caption{The solid line is the theoretical result for the critical
field as a function of temperature using the exchange parameters
introduced by Eq.\ (\ref{e31}). The dashed line is the result
obtained if using instead the exchange parameters of Oosawa {\it et
al.}\cite{Oosawa} The critical field is here defined to be the one at
which the paramagnetic phase becomes unstable. The experimental
points are those obtained when the field is applied in the $b$
direction by Oosawa {\it et al.}\cite{Oosawaheat} and Shindo and
Tanaka.\cite{Shindo}} \label{f2}
\end{figure}

Using the model defined here we have calculated the susceptibility as
a function of temperature. The result is compared with experiments in
Fig.\ \ref{f1}. The calculated critical field at which the
paramagnetic phase becomes unstable is compared with experiments in
Fig.\ \ref{f2}. The induced magnetic moment $m_z^{}$ at various
values of the field has been calculated as a function of $T$. These
results are shown in Fig.\ \ref{f3}. This figure also includes
results obtained in the ordered phase, which is considered in the
following section.

\section{Excitations in the antiferromagnetic phase}\label{sec:level3}
The paramagnetic phase becomes unstable when the energy of the lowest
excitation vanishes. The lowest energy mode is the one with the
energy $E_{\bm q}^-$ at ${\bm q}={\bm Q}=(001)$, and below the
transition the expectation value of the dimer-spin variable $\bar{\bm
S}_i^{}$ becomes non-zero. The ordering is antiferromagnetic in the
sense that $\langle\bar{\bm S}_i^{}\rangle$ have opposite signs on
the two chain sublattices. The antiferromagnetic ordering may be
transformed to the uniform one by an interchange of the two spins in
the definition of $\bar{\bm S}_i^{}$ for the dimers belonging to, for
instance, the 3-4 sublattices. The only effect of this transformation
is that the interchain coupling between the 1-2 and 3-4 dimers
changes sign, i.e.\ the $\pm$ in front of the second term of $J({\bm
q})$ in Eq.\ (\ref{e3}) is being replaced by $\mp$, and $J({\bm Q})$
and $J({\bm 0})$ are being interchanged within the extended zone
scheme. That the ordering is antiferromagnetic instead of being
uniform does not introduce any further complications.
$\langle\bar{\bm S}_i^{}\rangle$ is perpendicular to the
$z$-direction of the field, but its direction within the $x$--$y$
plane is arbitrary (as long as any anisotropy is neglected). For
convenience we shall make the choice that the ordered moment is along
the $x$ axis, and we define
\begin{equation}
\label{e34}m_{xy}^{}=g\mu_B^{}\frac{\langle\bar{S}_x\rangle}{2}\,,
\qquad\langle\bar{S}_x\rangle=\frac{1}{N}\sum_i\langle
\bar{S}_{ix}^{}\rangle e^{i{\bm Q}\cdot{\bm R}_i}\,.
\end{equation}

\begin{figure}[t]
\includegraphics[width=0.95\linewidth]{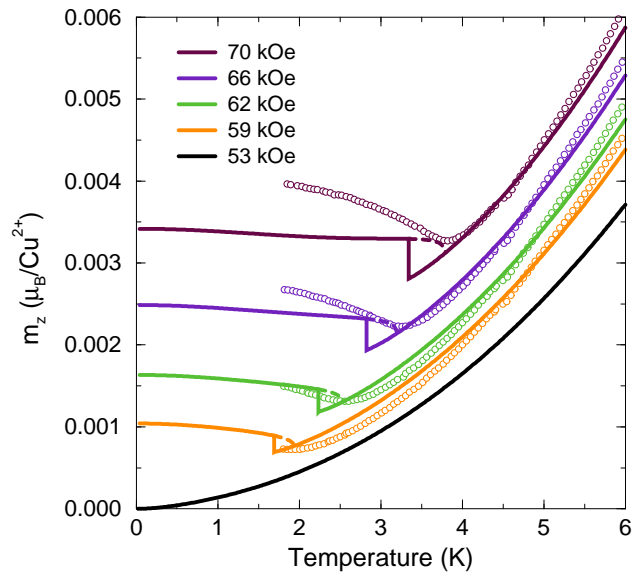}
\caption{(Color online) The parallel magnetic moment per Cu ion,
$m_z^{}$, as a function of temperature calculated at various values
of the field applied along the $b$ direction. In the case of $H=53$
kOe the system is predicted to stay disordered all the way to zero
temperature. The experimental points are a selection of those
obtained by Oosawa {\it et al.}\cite{Oosa99,Nikuni}}\label{f3}
\end{figure}

When $\langle\bar{S}_x\rangle$ is non-zero, the MF Hamiltonian for
the ``non-interacting'' $i$th dimer becomes
\begin{eqnarray}
\label{e35} {\cal
H}_{\text{MF}}^i=\!\!\!\!&&(\Delta-h)\,a_{11}^i+\Delta\,a_{22}^i+
(\Delta+h)\,a_{33}^i\nonumber\\
&&-J({\bm Q})\frac{1}{\sqrt{2}}\left[a_{30}^i-a_{10}^i+
a_{03}^i-a_{01}^i\right]\langle\bar{S}_x\rangle
\end{eqnarray}
in terms of the standard basis operators of the paramagnetic system.
We shall continue to label the eigenstates by $|\mu\rangle$, where
$\mu=0,1,2,3$, and the ground state of this MF Hamiltonian may then
be written
\begin{eqnarray}
\label{e36} |0\rangle=\cos\theta\,|00\rangle-\sin\theta\left[
\cos\left(\alpha-{\textstyle\frac{\pi}{4}}\right)\,|1+1\rangle
\quad\nonumber\right.\\
+\left.\sin\left(\alpha-{\textstyle\frac{\pi}{4}}\right)\,|1-1\rangle\right]\,.
\end{eqnarray}
The two angles $\theta$ and $\alpha$ minimize $\langle0|{\cal
H}_{\text{MF}}^i|0\rangle$, or may be determined by demanding the
off-diagonal terms of the MF Hamiltonian to vanish, and we get
\begin{equation}
\label{e37} \tan\alpha=\frac{h\cos2\theta}{\Delta\cos^2\theta-
h\sin^2\theta\sin2\alpha}
\end{equation}
and
\begin{equation}
\label{e38}\sin2\theta=\frac{2J({\bm Q})\cos\alpha}
{\Delta\cos2\alpha}
\left(\cos2\theta+2\sin^2\theta\sin^2\alpha\right)\langle\bar{S}_x^{}\rangle\,.
\end{equation}
By calculating the state vectors of the excited MF levels to order
$\theta^2$, we find that the order parameter is
\begin{equation}
\label{e39}
\langle\bar{S}_x^{}\rangle=\frac{n_{01}+n_{03}}{2}\sin2\theta\cos\alpha
-n_{13}^{}\sin\theta\sin\alpha+{\cal O}(\theta^3)\,,
\end{equation}
where the higher-order terms ${\cal O}(\theta^3)$ vanish if
$n_0^{}=1$, and, likewise, the ferromagnetic component is
\begin{equation}
\label{e40}\langle
S_z^{}\rangle=n_{13}^{}\cos\theta+\frac{n_{01}+n_{03}}{2}\sin^2\theta
\sin2\alpha+{\cal O}(\theta^4)\,.
\end{equation}
By $h_c^{}$ we denote the critical field at which these equations
have a non-zero solution for $\langle\bar{S}_x^{}\rangle$ in the
limit of $\theta\to0$, and this field is found to be determined by
\begin{equation}
\label{e41} 1-\left(\frac{h_c^{}}{\Delta}\right)^2=\frac{J({\bm
Q})}{\Delta}
\left(n_{01}^{}+n_{03}^{}-\frac{h_c^{}}{\Delta}\,n_{13}^{}\right)\,.
\end{equation}
If we replace $J({\bm Q})$ by $J_{xy}^{}({\bm Q})$ and $\Delta$ by
$\Delta_1^{}$ this condition is the same as that derived from the
requirement that the energy of the lowest paramagnetic excitation
$E_{\bm Q}^-$, within the self-consistent RPA, should vanish at the
transition. The results above are more general but coincide with
those derived in the zero-temperature mean-field theory of Matsumoto
{\it et al.}\cite{Matsumoto} [Their angle $\phi$ corresponds to our
$\frac{\pi}{4}-\alpha$. In our notation $h=h_0^{}+J_F^{}({\bm
0})\langle S_z^{}\rangle$, whereas in their notation $h$ in Eq.\
(\ref{e37}) should read $h-J({\bm Q})\langle S_z^{}\rangle$
corresponding to the replacement of $h_0^{}$ by $h$ with their
implicit assumption that the ferromagnetic interaction $J_F^{}({\bm
0})$ is equal to $-J({\bm Q})$].

When the MF Hamiltonian of the antiferromagnetic phase has been
diagonalized we may proceed as in the paramagnetic case for
calculating the correlation functions. The positions of the four
different levels and the matrix elements of $\bar{\bm S}_i^{}$ may be
calculated analytically if only terms to leading order in $\theta^2$
are included. We are not going to present these results, since in the
final calculations we chose the more accurate approach of
diagonalizing the MF Hamiltonian numerically. In terms of the
standard basis operators of the final MF Hamiltonian, $\bar{S}_{jx}$
is now $[m_3^{}(a_{30}^{j}+a_{03}^{j})
-m_1^{}(a_{10}^{j}+a_{01}^{j})] /\sqrt{2}$, where $m_1^{}$ and
$m_3^{}$ are different from 1 and from each other. Here we neglect
the extra complication that the matrix elements between the excited
states, as for instance $\langle 1|\bar{S}_{jx}|3\rangle\propto
\theta$, become non-zero. These excited state contributions to the
correlation functions get multiplied by $n_{13}^{}$, hence they are
unimportant not only when the ordered moment is small, but in most of
the regime where the RPA modifications of the MF behavior are of
importance.

The final Hamiltonian may be written in the same way as in the
disordered case, Eq.\ (\ref{e6}). The positions of the three excited
levels are being shifted and $J(ij)$ is being multiplied by different
factors depending on which operator product is considered, and,
finally, the remaining off-diagonal products, $a_{01}^ia_{01}^j$,
$a_{01}^ia_{30}^j$ etc., now appear. The new off-diagonal
contributions are all of order $\theta^2$ and affect the diagonal
correlation functions only to order $\theta^4$. Hence, to leading
order these extra contributions may be neglected. In this case the
matrix equations once again decouple into 3 sets of $2\times2$
equations, which may be solved analytically. The result for the
population numbers is the same as that given by Eqs.\
(\ref{e20})-(\ref{e23}), except that $\Delta$ and $\Delta\pm h$ are
being replaced by the energies of the three corresponding MF levels
and that $J_1^{}$, $J_2^{}$, and $J_3^{}$ are being multiplied by
matrix-element factors, which are slightly different from 1 and from
each other. The equivalence implies that the modifications of the RPA
correlation functions may be calculated as in the paramagnetic case,
and, for instance, $a_{xy}^{}$ is still determined by Eq.\
(\ref{e28}) except that the expressions for $E_{\bm q}^{\pm}$ are
being modified. We used this approximation, valid in the limit of
$\theta^2$ being small, for calculating the magnetization curves. The
results were close to those shown in Fig.\ \ref{f3}, however, in the
final calculations we included the higher-order modifications.

\begin{figure}[t]
\includegraphics[width=0.99\linewidth]{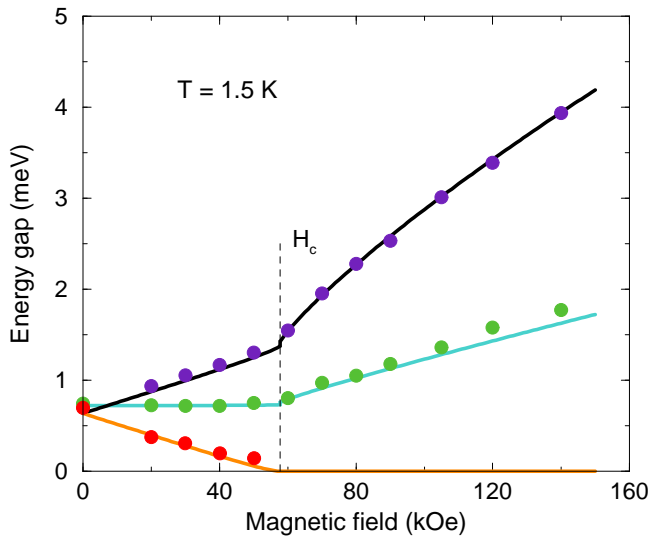}
\caption{(Color online) The minimum energies of the three different
dimer excitations as functions of field at 1.5 K. The calculated
results (the solid lines) are compared with the experimental results
of R\"uegg {\it et al.}\cite{Ruegg,RueggN}} \label{f4}
\end{figure}

For a given set of population numbers the MF Hamiltonian was
diagonalized numerically, determining all possible matrix elements
and the four energy levels. The knowledge of the population numbers
and the matrix elements is used for determining the two expectation
values $\langle \bar{S}_x^{}\rangle$ and $\langle S_z^{}\rangle$, and
for constructing the total Hamiltonian expressed in terms of the
standard basis operators. When the interactions between the excited
states are neglected, the equations of motion lead to a $4\times4$
set of matrix equations for the $xy$ part and a $2\times2$ set for
the longitudinal Green functions. The two sets of equations were
inverted analytically (utilizing Mathematica for handling the set of
$4\times4$ equations). In this way we derived an explicit expression
for the correlation function matrix $\bar{\bar{A}}({\bm q}^{})$. The
averaged values of the diagonal components were used for calculating
the population numbers as in the paramagnetic case. The
renormalization parameter $a_{xy}^{}$ and $a_z^{}$ are determined by,
respectively, $\overline{A}_{16}^{}=0$ and $\overline{A}_{34}^{}=0$.
The condition $\overline{A}_{16}^{}=0$ also implies that
$\overline{A}_{25}^{}=0$, but not necessarily that the new
off-diagonal components vanish. We have neglected the possibility
that the renormalization of the additional off-diagonal exchange
terms might be different, since the new terms are small whenever the
renormalization effects are important. Except for the matrix-element
modification of the exchange terms we use the same approximate
expression, Eq.\ (\ref{e27}), for the renormalization parameter
$b_{xy}^{}(\omega)$, and similarly for $b_{z}^{}(\omega)$. When the
renormalization parameters have been determined we may calculate the
renormalized value of $J({\bm Q})$ in the MF Hamiltonian (\ref{e35}),
which is being replaced by $J_{xy}^{}({\bm Q})$ determined from Eq.\
(\ref{e25}). Similarly $\Delta\pm h$ and $\Delta$ in Eq.\ (\ref{e35})
are replaced by, respectively, $\Delta_1^{}\pm h$ and $\Delta_2^{}$
given by Eq.\ (\ref{e30}). The MF Hamiltonian is then consistent with
the renormalized RPA expressions, and the whole procedure has been
carried out in a self-consistent manner, so that the population
numbers assumed as a start are the same as those derived.

\begin{figure}[t]
\includegraphics[width=0.99\linewidth]{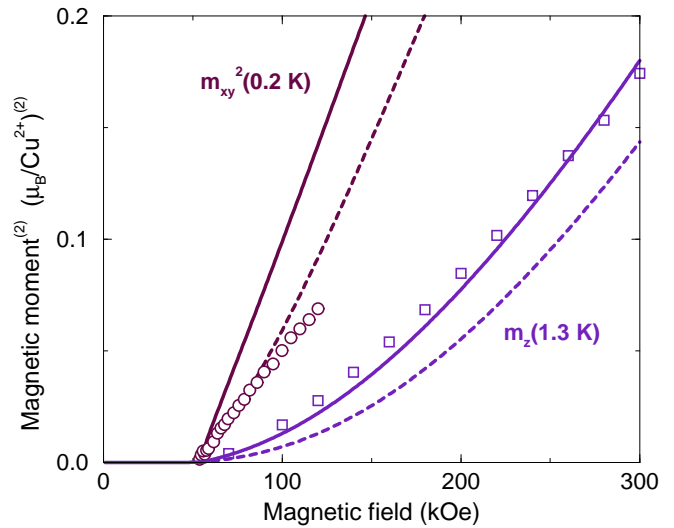}
\caption{(Color online) The squares are the experimental results for
the field dependence of the parallel moment $m_z^{}$ at 1.3 K and are
the data of Tatani {\it et al.}\cite{Tatani} presented in Ref.\
\onlinecite{Matsumoto}. The open circles are the neutron diffraction
results for the square of the ordered antiferromagnetic moment
$m_{xy}^{2}$ obtained by Tanaka {\it et al.}\cite{Tanaka}\ with the
field along the $b$ direction at 0.2 K. The solid lines show the
corresponding theoretical predictions, and the dashed ones are the
results of using the MF approximation.}\label{f5}
\end{figure}

This theory was used for calculating the properties of the dimer
system in the ordered phase. The temperature dependence of the
parallel moment in a constant applied field is shown in Fig.\
\ref{f3}. In Fig.\ \ref{f4} we show the calculated energies for the
three modes at $(001)$ as functions of field at 1.5 K compared with
the neutron-scattering results of R\"uegg {\it et
al.}\cite{Ruegg,RueggN} The experimental results indicate that
$\Delta$ for the longitudinal mode, the energy of which is nearly
unaffected by the field in the paramagnetic phase, is slightly larger
than for the two other modes, and we have accounted for this effect
by adding 0.03 meV to $\Delta_2^{}$ in Eq.\ (\ref{e30}). This
modification does not affect the $xy$-polarized modes, and as
discussed by Matsumoto {\it et al.},\cite{Matsumoto} the
lowest-energy $xy$-polarized mode becomes the Goldstone mode in the
ordered phase, the energy of which depends linearly on $|{\bm q}-{\bm
Q}|$ and is zero at the ordering wave vector.\cite{RueggN} The
spin-resonance experiments of Glazkov {\it et al.}\cite{Kolezhuk}
indicate that the Goldstone mode develops an energy gap for fields
larger than the critical one corresponding to the presence of a small
anisotropy within the $x$--$y$ plane of the same order of magnitude
as the anisotropy considered above. The two anisotropy terms are
unimportant for the renormalization effects and are neglected
elsewhere in the present calculations. The final figure, Fig.\
\ref{f5}, shows the field dependencies of the squared primary order
parameter $m_{xy}^{2}$ and of the parallel magnetization $m_z^{}$ in
the zero-temperature limit. The comparison with experiments shows
that the self-consistent RPA accounts reasonably well for the field
dependence of $m_z^{}$, whereas the order parameter is calculated to
increase rather faster with field than observed. The corresponding MF
model, on the other hand, underestimates the value of $m_z^{}$, but
predicts an order parameter which is close to the one observed.

\section{Discussion}\label{sec:level4}
The present dimer system is unique because it clearly exhibits the
importance of quantum fluctuations. The system is close to a quantum
critical point, and a zero-temperature phase transition may be
achieved either by the application of a modest magnetic field of 54
kOe\cite{Oosa99} or a hydrostatic pressure of 1.1
kbar.\cite{Ruegg2,Ruegg08} Here we have considered the case where the
transition is approached by applying a magnetic field at ambient
pressure. The field removes the degeneracy of the $S=1$ triplet
states of the dimers, and the collective singlet-triplet excitations
separate into a longitudinal, $z$-polarized wave and two transverse
modes. The collective transverse modes are linear combinations of
propagating modes due to transitions between the ground state and the
lowest and the highest excited states of the single dimers. In the
paramagnetic phase, the two transverse modes are subject to the same
renormalization effects, because the rigid energy shift of the
excitations, $E_{\bm q}^+-E_{\bm q}^-=2h$, does not influence the
quantum fluctuations and $n_{13}^{}=0$ at $T=0$. At non-zero
temperatures, the thermal fluctuations imply that $n_{13}^{}$ is
non-zero, but the renormalized exchange interaction $J_{xy}^{}({\bm
q})$ is still the same for the upper and lower transverse modes in
the paramagnetic phase (within the present approximation scheme).

\begin{figure}[t]
\includegraphics[width=0.99\linewidth]{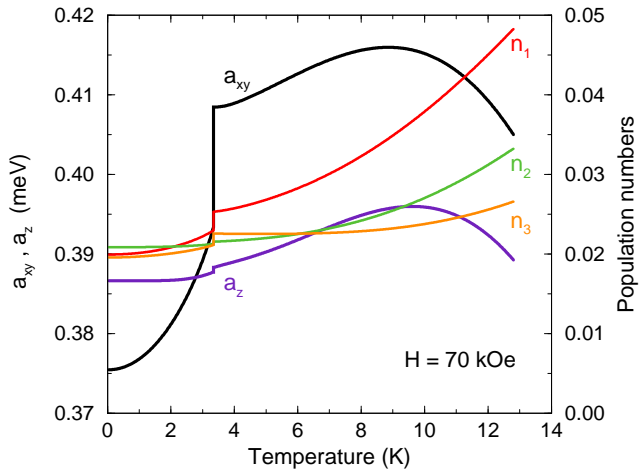}
\caption{(Color online) The renormalization parameters $a_{xy}$ and
$a_z^{}$ (left scale) and the population numbers (right scale)
calculated as functions of temperature at an applied field of 70 kOe,
where the transition occurs at 3.34 K.}\label{f6}
\end{figure}

The RPA renormalization parameters are determined to be rather
substantial in the zero-temperature limit at zero field, because the
system is close to the critical point. The occupation number of the
dimer ground state is calculated to be $n_0^0=0.935$ and
$a_0^{}=0.393$ meV. The constant reduction of the exchange
interaction is $(1+\eta_{xy}^0)a_0^{}=0.472$ meV, which is about 17\%
of the maximum value of the effective exchange interaction
$J_{\text{eff}}^{}({\bm Q})=2.8$ meV. The effective singlet-triplet
splitting is about $\Delta+a_0^{}=5.26$ meV in the temperature range
of the maximum in the susceptibility. As indicated by the comparison
in Fig.\ \ref{f1}, this is in good agreement with that derived from
the experimental data.\cite{Oosa99,Raffaele} Hence, the theory is
able to account for the difference found experimentally between the
(effective) energy gap of 5.68 meV, derived from the excitation
spectrum in the $T=0$ limit,\cite{Oosawa} and the smaller value of
the gap determined from the susceptibility measurements. At zero
temperature the renormalization parameters are independent of the
field as long as it stays smaller than the critical one. At non-zero
temperature the lower branch $E_{\bm q}^-$ is much more easily
populated than the corresponding MF level. For comparison, the bulk
magnetization at 53 kOe predicted by the corresponding MF model is
about a factor of 50 times smaller at 6 K. At a constant non-zero
temperature $E_{\bm q}^-$ decreases, and the thermal population of
the $E_{\bm q}^-$-branch increases, when the field approaches the
critical one. This also implies that the renormalization parameter
$a_{xy}^{}$ increases with increasing field. When the field becomes
larger than the critical one, $a_{xy}^{}$ is reduced as the field is
further increased. This reduction is so large that it is able to
stabilize the ordered moment also at fields smaller than the critical
one, i.e.\ the transition is so strongly modified that it becomes a
first order one. This behavior of the renormalization parameters at
the phase transition is illustrated in Fig.\ \ref{f6}. The unphysical
enhancement of the renormalization effects near a phase transition is
a quite general feature of the self-consistent version of RPA. In the
close neighborhood of the critical point the renormalization effects
show a pronounced sensitivity to small modifications of the model,
and the rather good agreement between theory and experiments obtained
here for the critical field and for the parallel magnetization, see
Figs.\ \ref{f2} and \ref{f3}, is somewhat fortuitous. Even the minor
change introduced by using the exchange parameters derived by Oosawa
{\it et al.}\cite{Oosawa} rather than those defined by Eq.\
(\ref{e31}) leads to a relative increase of $a_{xy}^{}$ by about 10\%
and to corresponding changes of the magnetization curves and the
critical field (see Fig.\ \ref{f2}).

Focusing our attention on the zero-temperature limit, then
$n_{13}^{}$ is zero when the field is smaller than the critical one,
but becomes non-zero in the ordered phase. As long as the ordered
moment is small, the contributions due to $n_{13}^{}$ may be
neglected and the equations determining $\langle\bar{S}_x^{}\rangle$
and $\langle S_z^{}\rangle$, Eqs.\ (\ref{e39}) and (\ref{e40}),
predict
\begin{equation}
\label{e42}\langle
S_z^{}\rangle=\frac{h}{\Delta(n_{01}+n_{03})}\langle\bar{S}_x^{}\rangle^2\,,
\end{equation}
when terms of the order of $\theta^4$ and $\alpha^2\theta^2$ are
omitted. The equation is only weakly influenced by the
renormalization effects since $n_{01}^{}+n_{03}^{}\approx 2$ at
$T=0$. Nevertheless, the experimental low-temperature results are far
from obeying this relationship, which circumstance makes it difficult
to reproduce the field dependencies of the two magnetization
components simultaneously, as illustrated by Fig.\ \ref{f5}.

In their modelling of the excitations by bosons, Nikuni {\it et
al.}\cite{Nikuni} find that $\langle
S_z^{}\rangle\simeq\frac{1}{2}\langle\bar{S}_x^{}\rangle^2$ at $T=0$
[in this expression we have neglected the small difference
$\tilde{n}$, between $n=\langle S_z^{}\rangle$ and $n_c^{}$, derived
by Nikuni {\it et al.}, which approximation corresponds to a
replacement of $n_{01}+n_{03}$ in Eq.\ (\ref{e42}) by 2]. Hence, the
theory of Nikuni {\it et al.}\ does not include the factor $h/\Delta$
appearing in our relation between $\langle S_z^{}\rangle$ and
$\langle\bar{S}_x^{}\rangle^2$. If this factor is included, the
results for $m_{xy}^{}$ of Nikuni {\it et al.}, i.e.\ $m_\perp^{}$ in
their Fig.\ 4, should be multiplied by $\sqrt{\Delta/h}\simeq 2.6$
leading to a slope of $m_{xy}^2$ with respect to field, which is
nearly twice the one derived by the present theory, i.e.\ a factor of
3-4 larger than the experimental one (see Fig.\ \ref{f5}). In the
paramagnetic phase, the lowest excited state of a single dimer is the
$|1+1\rangle$ state with $S_z^{}=1$, and Nikuni {\it et al.}\ are
assuming that this state is the one determining the wave functions of
the lowest lying mode of collective excitations, and hence the one
which defines the condensate in the ordered phase. This corresponds
to assuming $\alpha=\pi/4$ in our Eq.\ (\ref{e36}). However, as also
stressed by Matsumoto {\it et al.},\cite{Matsumoto} it is crucial to
include the presence of the $|1-1\rangle$ level in order to get a
consistent description of the excitations and of the condensate. In
the paramagnetic phase, the matrix element of $\bar{S}_x^{}$ between
the ground state $|00\rangle$ and the lowest excited state
$|1+1\rangle$ is numerically the same as its matrix element between
$|00\rangle$ and $|1-1\rangle$. The same applies to $\bar{S}_y^{}$,
and this means that the collective transverse excitations transmitted
via these two operators are mixed $S_z^{}=\pm1$ excitations. Being
proportional to $J({\bm q})$, the degree of mixing depends on the
wave vector and is at its maximum at the ordering wave vector. In
correspondence to this, the MF ground state in the ordered phase,
Eq.\ (\ref{e36}), involves $|1-1\rangle$ as well as $|1+1\rangle$.
The two states are of equal importance in the limit of zero field,
and the relative weight of the two states is shifted from 1 in the
presence of a field as described by the angle $\alpha\simeq
h/\Delta$. The two components depend differently on $\alpha$, e.g.\
$\langle S_z^{}\rangle=0$ whereas $\langle\bar{S}_x^{}\rangle$ has
its maximum at $\alpha=0$, and the factor $h/\Delta$ in Eq.\
(\ref{e42}) is a simple consequence of this difference.

The present self-consistent RPA theory accounts reasonably well for
the paramagnetic properties of the dimer system. Within the MF model,
the bulk susceptibility vanishes exponentially in the
zero-temperature limit, whereas the present RPA model predicts a
power law $m_z^{}/H\propto T^\phi$ with $\phi=1.8$ at $H=53$ kOe.
This is consistent with experiments and the RPA theory also predicts
the right critical field for the phase transition. In contrast to the
boson model of Eq.\ (\ref{e0}), the present theory does not rely on
any free parameters. Note, however, that if we use the effective
exchange parameters, which Oosawa {\it et al.}\ determined from their
measurements of the dimer excitation spectrum,\cite{Oosawa} the
critical field is derived to increase slightly faster with
temperature than observed, as shown in Fig.\ \ref{f2}. This minor
discrepancy was neutralized by a small adjustment of the density of
states of $J_{\text{eff}}^{}({\bm q})$. Actually, it would have been
a surprise if the self-consistent RPA theory had been able to predict
the right critical field without any adjustments. In all
circumstances, it is clear that the theory needs to be corrected due
to critical fluctuations, since the self-consistent RPA predicts the
phase transition to the antiferromagnetic phase to be of first order
in contradiction with experiment.

The dimer system has a number of unusual magnetic properties. The
most outstanding one is that the system is driven into the phase of
antiferromagnetic order by the application of a uniform field.
Another unusual property shown by the ordered phase is that the bulk
magnetization increases, when the temperature is lowered at a
constant field, as this happens in spite of the fact that the
``degrees of freedom'' are being reduced because of the accompanying
enhancement of the antiferromagnetic order parameter. This behavior
is not in accordance with the MF model, whereas the self-consistent
RPA theory accounts, at least, qualitatively for this observation.
The self-consistent theory of the ordered phase is complicated, and
we have been forced to neglect a number of effects. One of the
complications, which has not been mentioned above, is that the matrix
elements of ${\bm S}_i^{}$ between the ground state and the excited
states are no longer zero in the ordered state implying additional
modifications of all normal modes of the system (the two classes of
modes may mix because the Cu sites lack inversion symmetry). We may
also add that the non-zero values of the diagonal elements of
$\bar{S}_x^{}$ in the ordered phase effectively give rise to
additional contributions to $\Delta_1^{}$ and $\Delta_3^{}$.
Fortunately, these extra complications should be unimportant within
the regime of low temperatures and small order parameter, where the
theory is applied.

The approximations made for $b_{xy}^{}(\omega)$ in Eq.\ (\ref{e27})
are only acceptable, when the renormalization effects due to this
term are small. This term is handled in a more rigorous way by a
diagrammatic high-density expansion.\cite{Stinch} To first order in
$1/z$ (where $z$ is the number of interacting neighbors), the
diagrammatic theory may be formulated in a self-consistent,
``effective medium'' fashion, which is the equivalent of the present,
self-consistently improved version of RPA.\cite{JJ,JJz} This $1/z$
theory has been applied to the Ising systems HoF$_3$ and
LiHoF$_4$,\cite{HoF3,LiHoF4} and in these cases the phase transitions
are predicted to remain of second order. Therefore, we expect that
the use of the $1/z$ expansion theory, to first order in $1/z$, is
going to reproduce the self-consistent RPA results derived here for
the paramagnetic phase, and should lead to an improved description of
the ordered phase.

The present RPA theory establishes a classification of the
renormalization effects, which affect the properties of this
quantum-critical dimer system. The most important one is the constant
reduction of the exchange interaction by $(1+\eta_{xy})a_{xy}$. This
term is equivalent to that produced by an on-site repulsive
interaction $J_{xy}^{}(ii)$ and is here found to be determined
directly from the higher-order modifications of the RPA Green
functions. Our analysis of the spin model also predicts the presence
of other renormalization effects, such as the $\omega$-dependent
correction $b_{xy}^{}(\omega)$ to $J_{xy}^{}({\bm q})$ and the
increase of the effective splitting between the single-dimer energy
levels by $a_{xy}^{}$ or $a_z^{}$. The ${\bm q}$-independent
reduction of the exchange interaction has its parallel in the
phenomenological repulsive interaction, $v_0^{}$ in Eq.\ (\ref{e0}),
in the boson model of Nikuni {\it et al.},\cite{Nikuni} whereas the
two other renormalization effects have no counterpart in their
theory. A more problematic simplification made by Nikuni {\it et
al.}\ is their assumption that the low temperature properties of the
system are dominated by one type of bosons, whereas, in reality, the
system contains three different kinds, where those corresponding to
$S_z^{}=1$ and $S_z^{}=-1$ are mixed. The degree of mixing depends on
wave vector and on field, and is important for the characterization
of the bosons in the condensate.

One basic difficulty in the many-body theory of localized spin
systems is that the operators describing the dynamics of the single
spins are not bosonic but more complicated operators, as indicated by
Eq.\ (\ref{e10}). This complication is responsible for the need to
renormalize the simple RPA theory. Although the present
self-consistent theory includes the leading-order renormalization
effects, the comparison between theory and experiments within the
ordered phase of TlCuCl$_3$ is not satisfactory. The diagrammatic
$1/z$ theory,\cite{Stinch,JJ,JJz} represents a more systematic
approach and should be able to give a more acceptable description of
the ordered phase. We expect, however, that the pronounced
experimental violation of the MF/RPA relation Eq.\ (\ref{e42}),
between the bulk magnetization and the ordered antiferromagnetic
moment, will remain a challenge to future theory.

\begin{acknowledgments}
We thank Kim Lefmann for stimulating discussions.
\end{acknowledgments}

\end{document}